\documentclass[reprint,prl,superscriptaddress]{revtex4-1}

\usepackage{graphicx}% Include figure files
\usepackage{dcolumn}% Align table columns on decimal point
\usepackage{bm}% bold math
\usepackage{amsmath, amsthm, amssymb}
\usepackage{xcolor}

\newcommand{\be}{\begin{equation}}
\newcommand{\ee}{\end{equation}}
\newcommand{\bea}{\begin{eqnarray}}
\newcommand{\eea}{\end{eqnarray}}

\newcommand{\bef}{\begin{figure}}
\newcommand{\enf}{\end{figure}}

\begin{document}
\draft
%*************************************************************************

%\title{Molecular Rydberg Laser without inversion}
%\title{Molecular Rydberg Laser 'without inversion'}
%\title{XUV lasing driven by strong laser field}
\title{XUV lasing during strong-field assisted transient absorption in molecules}

%*************************************************************************
\author{Timm Bredtmann}
\email{Timm.Bredtmann@mbi-berlin.de}

\affiliation{Max-Born-Institut, Max-Born-Strasse 2A, D-12489 Berlin, Germany}

\author{Szczepan Chelkowski}

\affiliation{Laboratoire de Chimie Th\'{e}orique, Facult\'{e} des Sciences, Universit\'{e} de Sherbrooke, Sherbrooke, Qu\'{e}bec, Canada J1K 2R1}

\author{Andr\'{e} D. Bandrauk}

\affiliation{Laboratoire de Chimie Th\'{e}orique, Facult\'{e} des Sciences, Universit\'{e} de Sherbrooke, Sherbrooke, Qu\'{e}bec, Canada J1K 2R1}

\author{Misha Ivanov}
\affiliation{Max-Born-Institut, Max-Born-Strasse 2A, D-12489 Berlin, Germany}

\affiliation{Department of Physics, Imperial College London, South Kensington Campus, SW7 2AZ, London, UK}

\date{\today}

\begin{abstract}

Using ab-initio non-Born-Oppenheimer simulations, we demonstrate amplification of XUV radiation
in a high-harmonic generation type process using the example of the hydrogen molecular ion.
A small fraction of the molecules is pumped to a dissociative Rydberg state
from which IR-assisted XUV amplification is observed.
We show that starting at sufficiently high IR driving field intensities the
ground state molecules become quasi-transparent for XUV radiation,
while due to stabilization gain from Rydberg states is maintained, thus leading to lasing from strongly driven
Rydberg states. Further increase of the IR intensity even leads to gain by initially
unexcited molecules, which are quickly excited by the 
driving IR pulse.

\end{abstract}
\maketitle

High-order harmonic generation (HHG) in gases is a table-top technique producing attosecond pulses from 
the vacuum ultraviolet to the soft x-ray region and beyond \cite{Popmintchev08062012},
opening, in particular, the field of attosecond science \cite{attosecondphysics}. Its challenge, 
however, is the low conversion efficiency. One option to address this challenge is to rely on phase matching, see e.g. \cite{Murnane_Science_1998,Murnane_Nature_2003,Murnane_Science_2003,zhang}. 
Alternatively, HHG may be enhanced at the single atom/molecule level by 
enhancing the first step of the HHG process -- ionization -- seeding it with single attosecond pulses 
or pulse trains, see e.g. \cite{bandrauk_xuv,takahashi_xuv,biegert_xuv,brizuela_xuv,lein_2014}. 

Here, we focus on the recombination step in HHG and explore the possibility of IR-assisted stimulated XUV recombination from Rydberg 
states, leading ideally to exponential growth of the 
XUV signal with the particle density, as opposed to quadratic growth in standard HHG (Figure 1(a)). 
While HHG is a coherent process, it begins as spontaneous: no incident field is
originally present at the HHG frequencies to stimulate the emission.
As the generated high-frequency light is
accumulated in the medium, it may start to stimulate
both emission and absorption at these new frequencies.
%This corresponds to self seeded amplification.
Whether these stimulated transitions would
lead to loss or self-seeded XUV light amplification 
depends on the dynamics of a quantum system. 
%Crucially, this system is dressed by the strong IR driving field,
Crucially, the strong IR driving field
alters and controls its response, see e.g. \cite{Harris_1998,Sokolov_1999,Sokolov_2000,Polovinkin_2011,Antonov_2013,Herrmann_2013} for
some striking examples of what such a control field can achieve. 

\begin{figure}[h]
  \centering
     \includegraphics[width=0.5\textwidth]{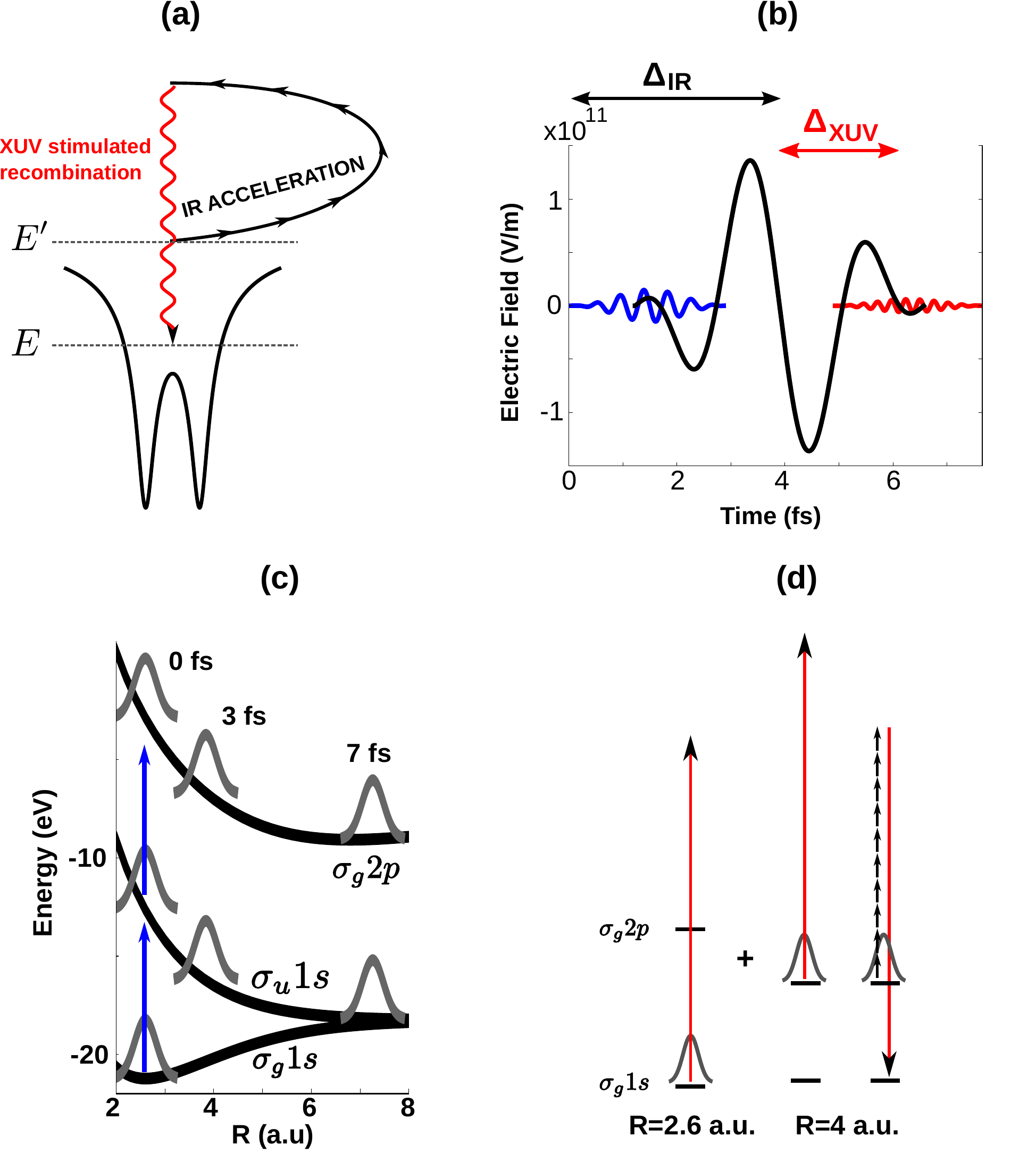}
  \caption{
Schematics of IR-assisted amplification of XUV light in the molecular ion H$_2^+$.
(a) In a HHG-type process, a strong IR pulse drives an electronic Rydberg state.
  Recombination is stimulated by the weak XUV seed present in the medium.
(b) Combination of VUV pump pulse (blue), IR driving field (black) and XUV probe pulse (red) used in the simulations.
(c) A fraction of the molecules is pumped via a two photon transition to a dissociative electronic Rydberg state.
  Timing of the IR driving field allows the control of XUV amplification.
(d) In the absence of the IR pulse, XUV absorption (red arrow) is possible both
  by the unexcited (left, internuclear distance R=2.6 a.u.) and the excited molecules (right, sketched for internuclear distance R=4 a.u.). The intense IR pulse can strongly attenuate XUV absorption while
assisting XUV recombination, reminiscent of 'lasing without inversion'.
  }
\end{figure}
If an effective population inversion is
created between the states during or after the interaction with the strong IR driving field,
light emitted at the corresponding transition frequencies can be amplified.
Parametric amplification might also be possible, with the energy transferred from
the driving field to the XUV field via nonlinear interaction with the medium. This mechanism was
suggested \cite{seres_NaturePhys_2010,seres_PRA_2012} as explanation of HHG
amplification reported in Ref. \cite{seres_NaturePhys_2010},
see also Ref. \cite{nersesov}.
Our ab-initio simulations include both
possibilities.

Excitations of Rydberg states during high harmonic generation occur e.g. due to frustrated tunnelling \cite{eichmann_tunneling,eichmann_tunneling2}, via field-induced Freeman 
resonances \cite{freeman_PhysRevLett.59.1092} or due to the lower harmonics of the IR radiation generated in the medium. It may play an important role during HHG in dense gases.
Here, however, we model Rydberg excitation as a separate step. 
This choice allows for a systematic analysis of the parameter ranges involved. 
Likewise, we split the generation of the primary harmonic seed light and its amplification into separate stages, 
i.e. the XUV seed is added explicitly in the simulations. 
This choice allows us to look at the transient absorption setup, where
an XUV probe can be moved relative to the dressing IR driver, leading 
to amplification of the XUV probe.
Moreover such splitting could also help one to exploit 
the full potential of XUV amplification \cite{Gallmann}. 

Specifically, we consider the molecular medium H$_2^+$.  
A short femtosecond pulse (Figure 1(b), blue) excites
a nuclear wavepacket from the electronic ground state $\sigma_{\rm{g}}$1s 
into the dissociative Rydberg state $\sigma_{\rm{g}}$2p (Figure 1(c)) via a two-photon 
transition, with the intermediate state $\sigma_{\rm{u}}{}$1s. 
 Next, a time-delayed combination of a strong IR driving field (Figure 1(b), black) and a weak XUV probe pulse (red), which stimulates non-resonant absorption or emission, interacts 
with both the ground state and the small fraction of excited dissociating molecules.
Short duration of the time-delayed probe allows one to think of a transient, time-dependent electronic
spectrum for the dissociating molecules.

Figure 1(d) sketches the key absorption and emission channels 
for the XUV probe pulse (red arrows) with and without the strong IR field (black arrows). 
In the absence of the IR pulse, XUV absorption  
is the only channel both for the unexcited (left) and the excited (right) molecules. 
We show that adding the IR field to the probe XUV attenuates absorption 
from the unexcited molecules, while IR-assisted XUV stimulated recombination of the excited molecules 
to lower states leads to amplification of the XUV pulse accompanied
by the absorption of many IR photons. The amplified windows in the XUV range depend on
the internuclear distance and may be controlled by timing of the pump pulse and the XUV--IR
pair.
 
%The possibility of parametric amplification from the ground state molecules will is investigated as well.

Two comments are in order:
First, the amplifier may operate over a wide tunable spectral range 
%since it is based on a non-resonant process and is thereby not limited 
%to specific atomic or molecular resonances.
Second, we find that even if the number of excited molecules is much less than the number of
unexcited ones, IR-assisted XUV amplification can still occur, reminding of 'lasing without inversion'
(see e.g. \cite{harris_PhysRevLett.62.1033,Scully21011994,Kocharovskaya1992175,Scully1997,1464-4266-2-3-201}).
%Second, even if the number of excited molecules is less than the number of
%unexcited ones, amplification at the transition energies characteristic of excited species, plus a number of IR photons,
%will still be possible.

We solve numerically the 3-body, Non-Born-Oppenheimer 
time-dependent Schr\"{o}dinger equation for the model molecule H$_{2}^{+}$ 
(atomic units, a.u., e=$\hbar$=$m_{{\rm el}}$=1 are used):
\begin{equation}
i\frac{\partial\Psi(z,R,t)}{\partial t}=H(z,R,t)\Psi(z,R,t),
\label{hamilt1}
\end{equation}
%including both electronic and nuclear degrees of freedom, 
where 
\begin{equation*}
H(z,R,t)=-\beta\frac{\partial^2}{\partial{}z^2} - \frac{1}{m_{\rm{p}}}\frac{\partial^2}{\partial{}R^2}+\frac{1}{R} + V_{\rm{C}}(z,R) + \kappa{}zE(t), 
\end{equation*}
with
\begin{equation*}
\beta = \frac{2m_{\rm{p}}+m_{\rm{e}}}{4m_{\rm{p}}m_{\rm{e}}}, \qquad \kappa=1+\frac{m_{\rm{e}}}{2m_{\rm{p}}+m_{\rm{e}}}
\end{equation*}
and the soft core potential
\begin{equation*}
V_{\rm{C}}(z,R)=\frac{-1}{\sqrt{(z-R/2)^2+1}}+\frac{-1}{\sqrt{(z+R/2)^2+1}},
\end{equation*}
is the three body Hamiltonian 
after separation of the center-of-mass motion. 
Both the nuclei and the electron are restricted to one dimension  
(see e.g. \cite{chelkowski_electron-nuclear_1998}). 
Here, z is the electron coordinate
%(with respect to the nuclear center of mass) 
and R is the internuclear
distance. 

%Initially ($t=0\,\rm{fs}$, Figure 1(b)), we consider a superposition of the ground electronic state 
%$|\sigma_{\rm{g}}1\rm{s}\rangle$ with a dissociating vibrational wavepacket on the excited electronic state 
%$|\sigma_{\rm{g}}2\rm{p}\rangle$, 
%$\Psi(t_0)=\Psi_{\rm{g}}+\Psi_{\rm{e}}= \chi_{\rm{g}}(R,t_0) \phi_{\rm{g}} +\chi_{\rm{e}}(R,t_0)\phi_{\rm{e}}$. 
%Here $\chi_{\rm{g}}(R,t)$ is the ground vibrational state on the 
%ground electronic potential energy surface, with the Born-Oppenheimer (BO) electronic wavefunction $\phi_{\rm{g}}$, 
%while  $\chi_{\rm{e}}(R,t)$ is the vibrational wavepacket 
%on the excited potential energy surface, with the BO electronic wavefunction $\phi_{\rm{e}}$.
%After a time-delay, both dissociating and unexcited molecules interact with the strong 800 nm few femtosecond IR field combined 
%with a weak, broad-band, XUV pulse, with total electric field $E(t)=E_{\rm{IR}}(t)+E_{\rm{XUV}}(t)$. The 
%XUV field  can stimulate high-frequency emission and absorption, both from the unexcited and excited molecules, probing the possibility of light
%amplification or loss. The center of the time-delayed IR pulse is held fixed at 3 fs while
%various time-delays $\tau_{\rm{XUV}}$ of the XUV seed with respect to the IR are
%considered (Figure 2(a)).
The electric field for each pulse is defined via the vector
potential $A(t)$ \cite{de_bohan_phase-dependent_1998}. Sine-squared functions are used for the
envelopes of the $A(t)$  for each pulse,  as in 
\cite{de_bohan_phase-dependent_1998,bredtmann_monitoring_2011}.
The carrier envelope phase $\phi$ for all pulses
(as defined in Ref. \cite{de_bohan_phase-dependent_1998}) is set to $\pi/2$, i.e., a sine pulse is
used. The total pulse durations are
$T_{\rm{IR}}=5.3\,\rm{fs}$ [2.7 fs full width at half maximum (FWHM)] for IR
and $T=2.7\,\rm{fs}$ [1 fs full width at half maximum (FWHM)] for both the VUV pump and the XUV seed, Figure 1(b).
While intensities of the 800 nm IR driving pulse and of the 134 nm VUV pump pulse are systematically varied, 
the intensity of the XUV seed with variable central frequency $\Omega_{\rm{XUV}}$ is set to
$I_{\rm{XUV}}=5\times{}10^{12}\,\rm{W/cm}^2$, which was tested to
yield stable numerical results while being in the linear response regime.

The total wavefunction of the system
is $\Psi(z,R,t)=\Psi_{\rm{IR}}(z,R,t)+\Delta{}\Psi_{\rm{XUV}}(z,R,t)$. Here
$\Psi_{\rm{IR}}(z,R,t)$ is the wavefunction of the IR dressed 
system, which also includes the effect of the pump pulse, and $\Delta{}\Psi_{\rm{XUV}}(z,R,t)$
is the perturbation due to the weak XUV probe. The induced dipole moment is
\begin{eqnarray*}
&&\langle\Psi(t)|\hat{d}|\Psi(t)\rangle =  \langle\Psi_{{\rm IR}}(t)|\hat{d}|\Psi_{{\rm IR}}(t)\rangle\\
 & & +\langle\Psi_{{\rm IR}}(t)|\hat{d}|\Delta{}\Psi_{{\rm XUV}}(t)\rangle+\langle\Delta{}\Psi_{{\rm XUV}}(t)|\hat{d}|\Psi_{{\rm IR}}(t)\rangle\\
 &  & +\langle\Delta\Psi_{{\rm XUV}}(t)|\hat{d}|\Delta\Psi_{{\rm XUV}}(t)\rangle,
\end{eqnarray*}
where $\hat{d}$ is the dipole operator. The first term on the right describes the non-linear response 
to the strong IR field alone, without any assistance from the XUV. Thus, it does 
not include any XUV stimulated processes 
and contains only spontaneous HHG-type emission during radiative recombination.
The second and third terms describe  
stimulated emission/absorption of XUV radiation by the IR dressed system, the subject of this paper.
The last term
depends quadratically on the weak XUV probe field and may thus be neglected for sufficiently low 
XUV intensities. 
For increased numerical stability, 
we use the dipole accelerations 
%$a(t)=-E(t)-\langle\Psi(t)|\frac{dV_{\rm{C}}}{dz}|\Psi(t)\rangle$ 
$a(t)=-\langle\Psi(t)|\big[ H,[H,z] \big] |\Psi(t)\rangle$ 
and 
%$a_{\rm{IR}}(t)=-E_{\rm{IR}}(t)-\langle\Psi_{\rm{IR}}(t)|\frac{dV_{\rm{C}}}{dz}|\Psi_{\rm{IR}}(t)\rangle$. 
$a_{\rm{IR}}(t)=-\langle\Psi_{\rm{IR}}(t)|\big[H,[H,z]\big]|\Psi_{\rm{IR}}(t)\rangle$. 
Hence, to identify stimulated absorption and emission of the XUV probe, we calculate the
frequency resolved linear response of the IR dressed system to the XUV probe pulse as
\begin{eqnarray}
	D_{\rm{XUV}}(\Omega)=\frac{1}{\Omega^2}\int dt e^{i\Omega t} (a(t) - a_{\rm{IR}}(t)),
% 	D_{\rm{XUV}}(\Omega)=\frac{1}{\Omega^2}\int dt e^{i\Omega t} M(t)\( \langle \Psi(t)|\hat a|\Psi(t)\rangle{} - \langle \Psi_{\rm{IR}}(t)|\hat a|\Psi_{\rm{IR}}(t)\rangle{} \),
	 \label{FT_dip}
\end{eqnarray}
thereby removing the contribution of the standard HHG-type emission. 
%unaltered by the weak XUV probe. 
The numerical propagation is carried out until the end of the XUV probe pulse.
%The mask  $M(t)=0.5*(1-\rm{cos}(2\pi{}(t-\tau_{XUV})/T_{\rm{IR}}))$ in 
%eq. \eqref{FT_dip} is introduced to reduce the numerical noise. 
%It is centered at the XUV field and extends over the entire duration of the IR pulse. 
%Thus, we effectively calculate the time-gated response within the temporal window 
%corresponding to the length of the IR pulse. One should keep in mind
%that, in the absence of the IR field, the temporal window introduces a slight XUV 
%carrier-frequency dependence into the purely linear response to the XUV field.
The XUV probe absorption/emission is related to the out-of-phase component 
of $D_{\rm{XUV}}(\Omega)$ with respect to the spectral amplitude of the XUV pulse, $E_{\rm{XUV}}(\Omega)$. 
The XUV probe absorption signal is  \cite{Mukamel}
\begin{equation}
S_{\rm{XUV}}(\Omega)\propto{}\frac{\rm{Im}(E^{\ast}_{\rm{XUV}}(\Omega)D_{\rm{XUV}}(\Omega))}{\int{}d\Omega{}|E_{\rm{XUV}}(\Omega)|{^2}}.
\end{equation}

\begin{figure}[t]
  \centering
    \includegraphics[width=0.5\textwidth]{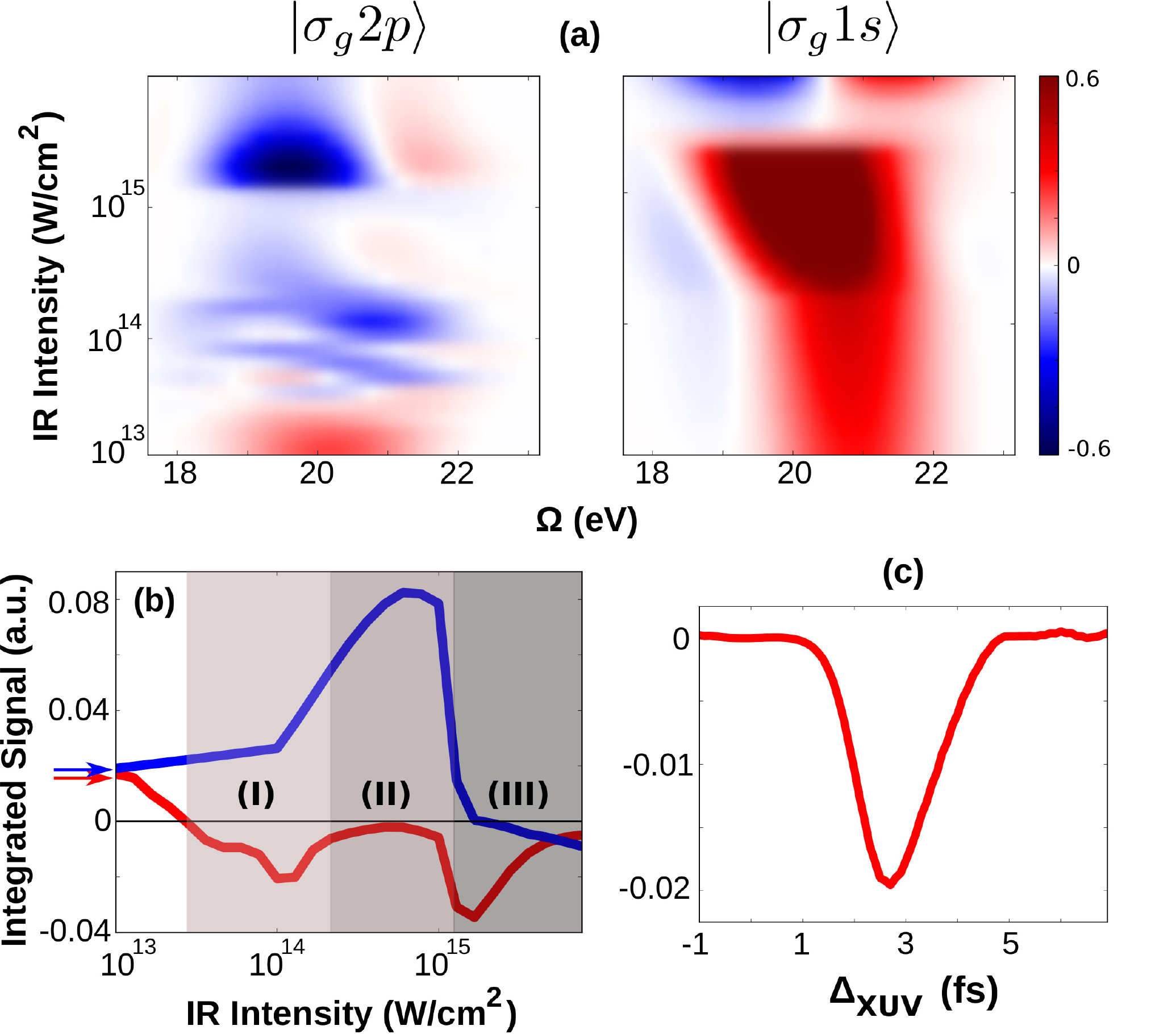}
  \caption{(a) XUV transient absorption spectra $S_{\rm{XUV}}(\Omega)$ from excited state molecules (left) and ground state molecules (right)
for variable IR intensities $I_{\rm{IR}}$ with $\Delta_{\rm{IR}}=3$\,fs, $\Delta_{\rm{XUV}}=2.4$\,fs (c.f. Figure 1(b)) and central XUV frequency $\Omega_{\rm{XUV}}$=20.4 eV. Here, we assume complete
population transfer to the excited state $\sigma_{\rm{g}}$2p at time zero. (b) Corresponding integrated signals, $\int{}S_{\rm{XUV}}(\Omega)$d$\Omega$, as function of the intensity of the IR driving field 
for excited state (red) and ground state (blue) molecules identifying three
amplification regimes. Values without the IR pulse are indicated by the horizontal arrows. (c) Dependence of integrated signals from excited state molecules on XUV-probe delay-time for 
I$_{\rm{IR}}$=5$\times$10$^{15}$\,W/cm$^2$.}
\end{figure}

\begin{figure}[t]
  \centering
    \includegraphics[width=0.4\textwidth]{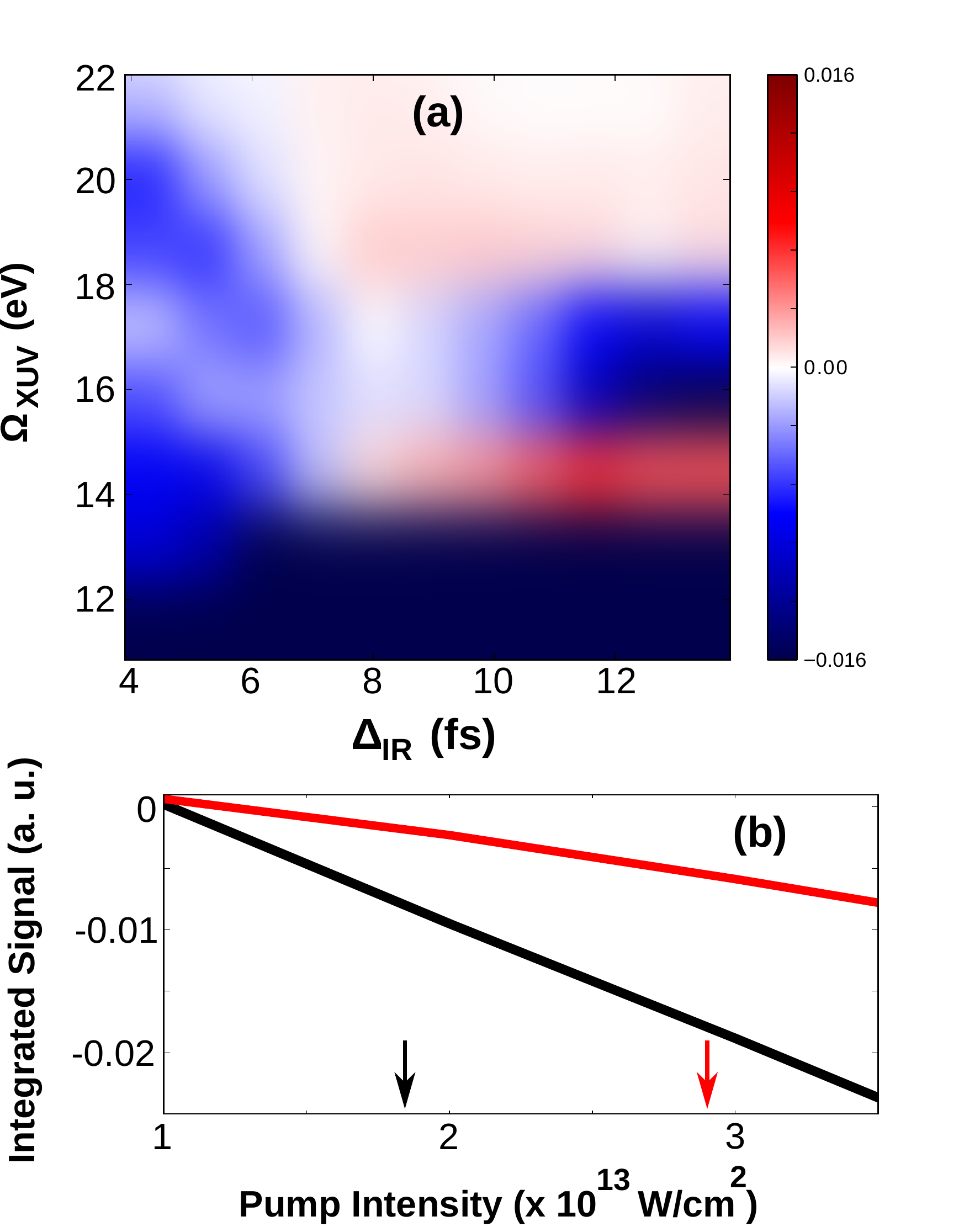}
  \caption{Full pump-probe simulations: (a) Integrated XUV transient absorption spectrum, $\int{}S_{\rm{XUV}}(\Omega)$d$\Omega$, as function of the
central frequency of the XUV seed, $\Omega_{\rm{XUV}}$, and the time delay of the IR driving field, $\Delta_{\rm{IR}}$.
The pump intensity is set to $I_{\rm{pump}}=3\times{}10^{13}$\,W/cm$^2$, exciting 16\% of the molecules to the dissociative Rydberg state $\sigma_{\rm{g}}$2p.
In all simulations, the intensity of the IR driving field
is set to $I_{\rm{IR}}=3\times{}10^{15}$\,W/cm$^2$ and the delay of the XUV seed is $\Delta_{\rm{XUV}}$=2.4 fs.
(b) Integrated signals as function of pump intensity, $I_{\rm{pump}}$, with central
XUV frequency $\Omega_{\rm{XUV}}=15.7$\,eV for large internuclear distances ($\Delta_{\rm{IR}}$=14.9 fs) (black)
and for $\Omega_{\rm{XUV}}=18.8$\,eV for short internuclear distances ($\Delta_{\rm{IR}}$=3.9 fs) (red). The vertical arrows indicate the pump intensities for which 
the overall gain matches the absorption from the unexcited molecules in the absence of the VUV pump and the IR pulse.}

\end{figure}

Figure 2(a) explores the possibility of light amplification from the excited molecules ($\sigma_{\rm{g}}$2p, left) and the reponse of the ground state molecules
($\sigma_{\rm{g}}$1s, right) for a highly non-resonant XUV probe-pulse for the intensity range I$_{\rm{IR}}$=1$\times$10$^{13}$\,W/cm$^2$ - 9$\times$10$^{15}$\,W/cm$^2$
of the IR driving field for time-delays $\Delta_{\rm{IR}}$=3 fs and $\Delta_{\rm{XUV}}$=2.4 fs, c.f. Figure 1(b). The central frequency of the XUV probe is set to
$\Omega_{\rm{XUV}}=\Omega_{\rm{ge}}+7\Omega_{\rm{IR}}$=20.4 eV, where $\Omega_{\rm{IR}}$ is the central frequency of the 800 nm IR pulse and $\Omega_{\rm{ge}}$=9.5 eV denotes the transition frequency
$|\sigma_{\rm{g}}2\rm{p}\rangle\rightarrow|\sigma_{\rm{g}}1\rm{s}\rangle$ for the employed delay-time. 
First, we assume complete population
transfer to the excited state $\sigma_{\rm{g}}$2p at time zero.
While there is only XUV absorption below I$_{\rm{IR}}$=5$\times$10$^{13}$\,W/cm$^2$, 
we identify three regimes for higher IR intensities, see Figure 2(b) 
which shows the integrated XUV transient absorption spectrum, 
$\int{S_{\rm{XUV}}}(\Omega)\rm{d}\Omega$, both for the excited state (red) and 
the ground state molecules (blue).
For comparison, the corresponding integrated signal strength without the IR driving field 
is indicated by the horizontal arrows. 

In regime (I), strong negative absorption from the excited state
molecules yields a first amplification maximum at I$_{\rm{IR}}$=1$\times$10$^{14}$\,W/cm$^2$. This gain from the excited state molecules is accompanied by comparable XUV light absorption
from the ground state molecules. 
In regime (II), rapid IR-induced depletion of the excited state reduces amplification from 
excited molecules and XUV absorption from the ground state molecules dominates. 
Starting with intensities I$_{\rm{IR}}$=2$\times$10$^{15}$\,W/cm$^2$ (regime (III)), 
one enters the stabilization regime for 
the excited states \cite{eichmann_tunneling,eichmann_tunneling2}, resulting
in striking XUV light amplification with maximum 
at I$_{\rm{IR}}$=3$\times$10$^{15}$\,W/cm$^2$, being even stronger 
than in regime (I). At the same time, the ground state
molecules become quasi-transparent due to rapid depletion.

For intensities beyond I$_{\rm{IR}}$=4$\times$10$^{15}$\,W/cm$^2$, lasing is even observed
from the molecules which were initially prepared in the electronic ground state. 
This is likely due to IR-induced transient population of Rydberg states through 
frustrated tunneling \cite{eichmann_tunneling,eichmann_tunneling2}. At these intensities, the gain from molecules 
initially in the electronic ground state adds to the stronger amplification of molecules
which were prepared in the Rydberg state $\sigma_{\rm{g}}$2p. 
This gain without nominal population inversion is stable over a large range of XUV-probe
time-delays $\Delta_{\rm{XUV}}$ (Figure 2(c)). Moreover, it is stable over a large range 
of XUV pulse durations (Supplementary Material), i.e. is robust with respect to possible XUV 
pulse reshaping during macroscopic light propagation through the molecular medium.

Figure 3 explores, for the full pump-probe scenario, the dependence of the integrated 
XUV transient absorption signal both on the delay of the IR driving field, $\Delta_{\rm{IR}}$, and
on the central frequency of the XUV seed, $\Omega_{\rm{XUV}}$. The pump intensity is set to I$_{\rm{pump}}$=3$\times$10$^{13}$\,W/cm$^2$, exciting 18\% of the ground state molecules
to electronic state $\sigma_{\rm{u}}$1s and 16\% to the dissociative Rydberg state $\sigma_{\rm{g}}$2p. 
The remaining parameters are identical to Figs. 2(a)+(b).
Even with 83\% of the molecules residing in the two lowest electronic states, tunable XUV light amplification is demonstrated. Control of the internuclear distance R through
the pump-probe time delay $\Delta_{\rm{IR}}$ allows the effective amplification of 
XUV pulses with central frequencies $\Omega_{\rm{XUV}}$ ranging from the 
pump carrier frequency ($\Omega_{\rm{XUV}}$=9.25 eV)
up to at least $\Omega_{\rm{XUV}}$=22 eV. 
Importantly, starting with time-delays $\Delta_{\rm{IR}}$=11 fs, the broad amplification windows 
centered around $\Omega_{\rm{XUV}}$=11 eV and $\Omega_{\rm{XUV}}$=17 eV
become independent of internuclear distance and hence no longer depend on 
the pump-probe synchronization. 

Finally, Figure 3(b) determines the pump intensities needed such that the overall gain in the full pump-probe simulations matches
the absorption from the ground state molecules in the absence of both the VUV pump and the IR pulse.
The results are indicated by the vertical arrows, black for XUV seeds centered at  
$\Omega_{\rm{XUV}}$=15.7 eV for long internuclear distances using delays $\Delta_{\rm{IR}}=14.9$ fs, and red for central frequency
$\Omega_{\rm{XUV}}$=18.8 eV for short internuclear distances using $\Delta_{\rm{IR}}=3.9$ fs.
The corresponding populations after the pump pulse in the state $\sigma_{\rm{g}}$2p are 8\% for $\Omega_{\rm{XUV}}$=15.7 eV (black) and 
15\% for $\Omega_{\rm{XUV}}$=18.8 eV (red).
%The overall gain can of course be controlled by increasing the pump intensity $I_{\rm{pump}}$, thereby 
%enhancing the number of molecules in the excited electronic state $\sigma_{\rm{g}}$2p.

In conclusion, we have shown that absorption of multiple IR photons from
the strong laser field can be accompanied by amplification of 
weak XUV light incident  
on the H$_2^+$ molecule. 
Amplification is enabled by excitation to a dissociative Rydberg state 
and controled by molecular dynamics. 
The XUV absorption from this excited Rydberg state 
is suppressed in favour of IR-assisted, stimulated XUV recombination into
the lower-lying bound states of the molecule. The process takes 
advantage of the stability of excited electronic states, with  
the strong-IR-driven wavefunction oscillating as a nearly free Kramers-Henneberger-like 
quasi-bound wavepacket \cite{Popov_2011,Felipe_PNAS_2011,Maria_NJPHYS_2013}. 
%including IR-induced excitations to higher lying Rydberg states. 

While for IR intensities below I$_{\rm{IR}}$=2$\times$10$^{15}$\,W/cm$^2$ 
strong absorption of the ground state molecules constitutes the limiting factor, we have 
established a high intensity regime (regime (III), Figure 2(b)), in which rapid 
ionization makes the overall response of the ground state molecules quasi-transparent. 
At the same time, considerable stabilization of Rydberg states yields overall gain even 
at small degrees of excitation.% around 5\%. 
%Normalized to the initial excitation, the gain is
%on the order of the absorption by the ground state molecules
%in the absence of the IR field.
 It is shown that exciting merely 10\%-15\% of the molecules suffices for the overall gain
to match the overall absorption of a sample of unexcited molecules in the absence of the IR pulse.

For intensities beyond I$_{\rm{IR}}$=4$\times$10$^{15}$\,W/cm$^2$ 
lasing of the initially unexciteted molecules is observed, adding to 
the considerably stronger gain from the molecules prepared by the pump pulse. 
This is most likely due to IR-induced transitions to Rydberg states 
through frustrated tunneling \cite{eichmann_tunneling,eichmann_tunneling2}.
%This amplifier is non-resonant and operates over wide spectral windows,
This amplifier is operates over wide spectral windows,
controled by the internuclear distance via the pump-probe time delay. 

Concerning the synchronization of the pulses, 
amplification occurs for a 
range of IR-XUV time-delays of about 3 fs, much longer than the XUV half-cycle. 
Thus, phase matching carrier oscillations 
of the XUV-IR pair within one XUV half-cycle is not required. 
%thereby replacing the challenging issue of phase matching by 
%the less stringent group velocity matching of the IR driver
%and the XUV seed. 
For the pump step, deviations of at least 1 fs are tolerated. Moreover, for 
specific but broad ranges of central XUV frequencies, e.g. around $\Omega_{\rm{XUV}}=17$\, eV, 
no pump-probe synchronization is neccessary for pump-probe delays $>$ 11 fs, 
until at least 50 fs. Hence, for such frequency windows, the external
addition of both the pump pulse and the XUV-seed may become 
obsolete in dense gases, resulting in self-seeded amplification, with 
excitation to Rydberg states originating either via multi-photon transitions induced by the VUV radiation 
generated in the medium, or through frustrated tunneling \cite{eichmann_tunneling,eichmann_tunneling2}.
%Our scheme has some 
%similarities to parametric wave-mixing schemes 
%in strongly driven systems \cite{Harris_1998,Sokolov_1999,Sokolov_2000,Polovinkin_2011,Antonov_2013,Herrmann_2013}. 
%However, none of these schemes uses
%ionization and dissociation-controled population inversion for excited molecular ensembles. 
% The efficiency of the mechanism decreases for large
% number of absorbed IR photons, when high-energy continuum states are populated. However, the efficiency 
% may be enhanced by increasing the intensity of the IR pulse until IR-induced ionization begin to dominate. 
%However, the efficiency of the mechanism appears to decrease rather quickly for large
%number of absorbed IR photons, when high-energy continuum states are populated.
% As yet, we have found no evidence of parametric amplification from the ground state molecules alone, which would
% correlate to the mechanism proposed in Refs. \cite{seres_NaturePhys_2010,seres_PRA_2012,nersesov}.

%The exciations investigated here may also have been created in the course of the experiment presented in Ref. \cite{seres_NaturePhys_2010}
%being the origin of the observed amplification.

\begin{acknowledgments}
T.B. and M.I. acknowledge partial financial suppot from US AirForce grant FA9550-12-1-0482.
M.I. acknowledges partial support of the EPSRC Programme Grant, and the Marie Curie CORINF network.
We thank S. Popruzhenko, O. Smirnova and S. Patchkovskii for stimulating discussions.
\end{acknowledgments}

\bibliography{amplification}

\clearpage

\section{Supplementary Material}

\begin{figure}[b]
  \centering
    \includegraphics[width=0.4\textwidth]{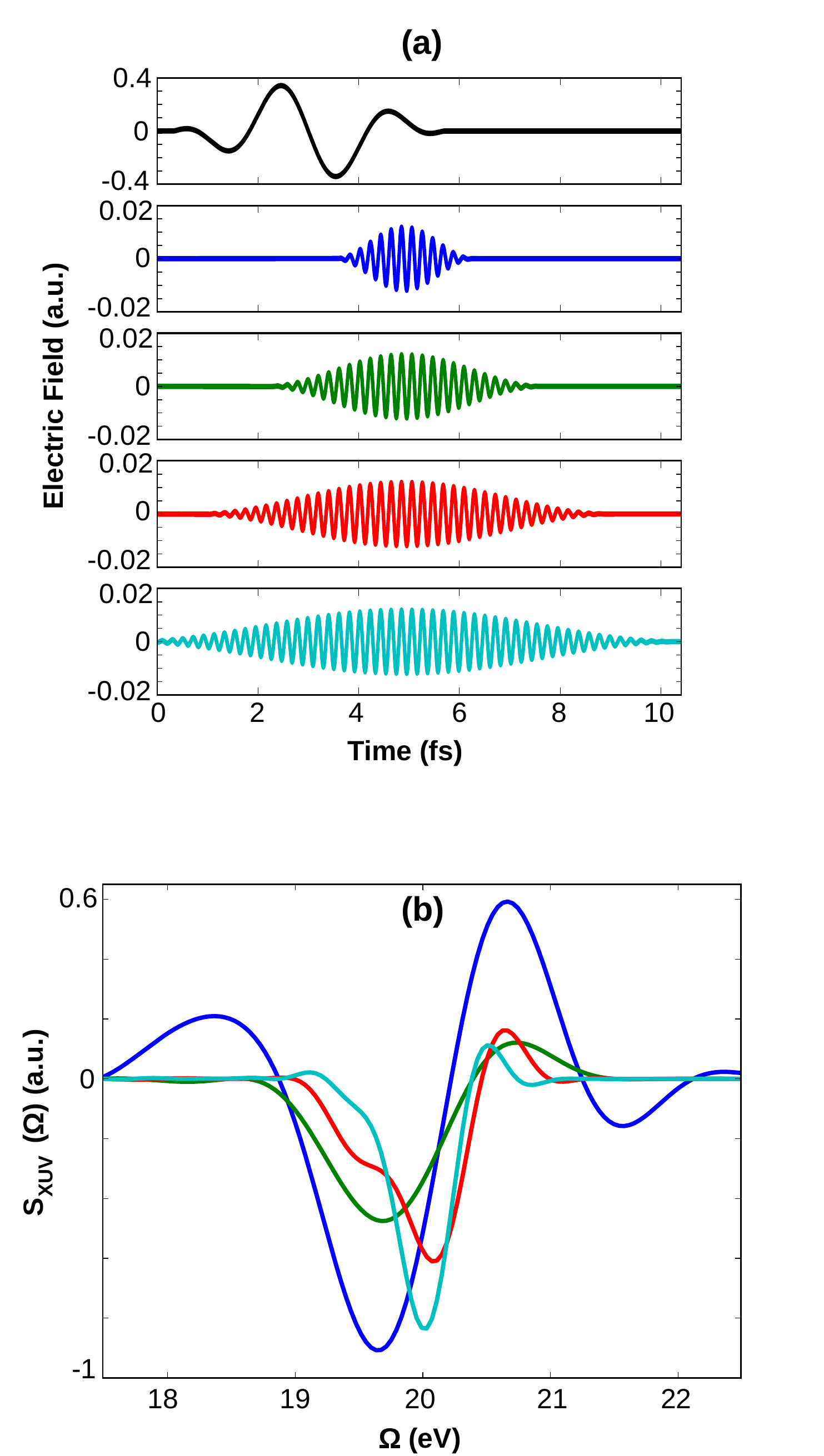}
  \caption{Influence of the XUV probe pulse duration, $T$, on the XUV transient absorption 
spectrum, $S_{\rm{XUV}}(\Omega)$, from excited $\sigma_{\rm{g}}$2p molecules,
for  $I_{\rm{IR}}=5\times{}10^{15}$ W/cm$^2$, $\Delta_{\rm{IR}}$=3 fs and $\Delta_{\rm{XUV}}=$1.9 fs and $\Omega_{\rm{XUV}}=$20 eV.  
 (a) From top to bottom: Electric field of the IR and XUV probe pulses with durations $T=2.75$ (fs) (blue, corresponding to the durations used in the main part of the text),
$T=5.5$ fs (green), $T=8.25$ fs (red), $T=11$ fs (cyan). (b) XUV transient absorption spectra, $S_{\rm{XUV}}(\Omega)$, for the XUV pulses from (a), shown in corresponding colors.
  }
\end{figure}

\end{document}